# Chess, Chance and Conspiracy

**Mark R. Segal**

*Abstract.* Chess and chance are seemingly strange bedfellows. Luck and/or randomness have no apparent role in move selection when the game is played at the highest levels. However, when competition is at the ultimate level, that of the World Chess Championship (WCC), chess and conspiracy are *not* strange bedfellows, there being a long and colorful history of accusations levied between participants. One such accusation, frequently repeated, was that all the games in the 1985 WCC (Karpov vs Kasparov) were fixed and prearranged move by move. That this claim was advanced by a former World Champion, Bobby Fischer, argues that it ought be investigated. That the only published, concrete basis for this claim consists of an observed run of particular moves, allows this investigation to be performed using probabilistic and statistical methods. In particular, we employ imbedded finite Markov chains to evaluate run statistic distributions. Further, we demonstrate how both chess computers and game data bases can be brought to bear on the problem.

*Key words and phrases:* Chess, data bases, distribution theory, Markov chains, run statistics, streaks.

> Chess *is* a game of luck. If you have a good opponent you have bad luck, and if you have a bad opponent you have good luck (Jacob Segal, age 7).

## 1. INTRODUCTION

Chess and chance are seemingly strange bedfellows. Many would contend that chess is the most logical/rational of all games, although such claims often get a rise from Go players. In any event, luck and/or randomness have no role in move selection when the game is played at the highest levels. However, when competition is at the ultimate level, that of the World Chess Championship (WCC), chess and conspiracy are *not* strange bedfellows. There is a long and colorful history of accusations, levied between participants, instances of which are showcased below. One such accusation, frequently repeated, was that all the games in the 1985 WCC were fixed and prearranged move by move. That this claim was advanced by a former World Champion, Robert (Bobby) Fischer, argues that it be investigated. Now, the logic that Fischer's chess playing credentials warrant such an investigation has been questioned (an Editor, Mark van der Laan, personal communications) in view of widespread perceptions concerning his diminished faculties (e.g., Krauthammer, 2005). However, it is worth noting that there are adherents to his allegations among the chess-playing community, including one former World Champion (Spassky, 1999) and one former women's World Champion (Z. Polgár in Polgár and Shutzman, 1997). Since the only published, concrete basis for Fischer's claim consists of an observed run of particular moves, we pursue such an investigation using probabilistic and statistical methods.

The paper is organized as follows. In the next subsection we provide a very brief history of the World

*Mark R. Segal is Professor, Department of Epidemiology and Biostatistics, and Director, Center for Bioinformatics and Molecular Biostatistics, University of California, San Francisco, California 94143-0560, USA e-mail: mark@biostat.ucsf.edu.*

Chess Championship post World War II. This provides context to the 1985 match. Conspiracy elements surrounding the 1978 match are highlighted to indicate the climate in which these contests are conducted. Section 2 focuses on Fischer's claim regarding move runs and several approaches for evaluating the significance thereof. In particular, we employ imbeddings in finite Markov chains that are especially suited for this purpose. Section 3 describes further analytic possibilities for run assessment that are afforded by use of powerful, chess-playing software, while Section 4 does likewise with respect to game data bases. Concluding discussion is presented in Section 5.

For the most part, only a very rudimentary understanding of chess is required for this paper. Rather than give definitions of pieces, legal moves and other rules, we defer such to the many expository treatments available both in print and online.

### 1.1 World Chess Championship: History and Controversy

Prior to 1948, the World Chess Championship was arranged at the behest of the reigning World Champion. This formulation allowed the Champion to select substandard opponents and avoid leading adversaries. However, the confluence of the death of the titleholder, World War II and the emergence of an international governing body, Fédération Internationale des Échecs (FIDE), led to the abandonment of such matches by invitation. In 1948, FIDE invited six leading players to participate in a round-robin tournament for the title of World Champion. Subsequently, up until 1990, the WCC was organized to run on a three-year cycle. The cycle started with the world's best chess players being seeded into one or more interzonal tournaments. The top finishers in the interzonal tournaments qualified for a series of elimination ("candidates") matches. The player who emerged victorious from the candidates matches met the reigning World Champion in a title match.

This period, which notably coincided with the Cold War, was marked by the dominance of Soviet players. From 1948 up to 1972 there were nine WCCs, each featuring a Soviet champion and challenger. However, in 1972, following a stunning surge through the interzonal tournament and candidates matches that included an unprecedented run of 19 consecutive wins, an American, Robert (Bobby) Fischer, emerged as the challenger to titleholder Boris Spassky. Fischer won convincingly, but forfeited the title to Anatoly Karpov in 1975.

Karpov's challenger in the 1978 WCC was Victor Korchnoi. Previously, in 1976, Korchnoi had defected from the Soviet Union and sought political asylum in the Netherlands. The USSR Chess Federation put pressure on FIDE to exclude Korchnoi from the candidates matches. FIDE did not buckle and, as fate would have it, Korchnoi defeated three Soviet players on his way to the title match with Karpov. Shortly before the final candidates match, Korchnoi was injured in a serious car accident. Then, during the WCC itself, a remarkable series of claims and counterclaims was exchanged, that showcases the tensions and controversies surrounding chess at this level. These included (i) Karpov demanding the dismantling of Korchnoi's chair to search for "extraneous objects or prohibited devices," (ii) Korchnoi wearing mirrored sunglasses to neutralize Karpov's habit of staring at his opponent, (iii) Korchnoi accusing Karpov of receiving move advice encoded by the color of the yogurts that he was given during the game, (iv) Karpov employing a parapsychologist, Dr. Zukhar, who sat in the audience, fixedly staring at Korchnoi, purportedly to disturb/hypnotize—this led to intense bickering with regard to Zukhar's seating position and Korchnoi placing his own hypnotist and two Ananda Marga sect members in the audience.

Ultimately, Karpov (barely) prevailed. He decisively beat Korchnoi in the subsequent 1981 WCC. Karpov's next challenger for the 1984 WCC was Gary Kasparov, a mere 21-year-old, who had dominated his three candidates matches en route to qualifying. Rules for the 1984 WCC were that the winner would be the first player to achieve six victories, there being no limit on the total number of games played. Karpov started convincingly, leading 4–0 after 9 games and 5–0 after 27. However, Kasparov won the 32nd game and, following another long string of draws, won both the 47th and 48th games. While Karpov still led 5–3, he had lost 10 kg (22 lbs), been hospitalized several times and was on the verge of collapse, his condition compounded by alleged frequent use of stimulants. Amid great controversy, the president of FIDE canceled the match, citing the failing health of the players due to the record breaking length and duration (6 months) of the contest. Provisions were made for a rematch to be held in 6 months time, with a fixed number (24) of games, Karpov retaining his title in the event of a tie. Kasparov won the 1985 rematch, becoming the youngest ever World Champion. He successfully, albeit narrowly, defended his title in three subsequent (1986, 1987, 1990) WCC matches against Karpov. It is this body of Karpov vs Kasparov games, and accusations surrounding them, that is the subject of our further analysis.

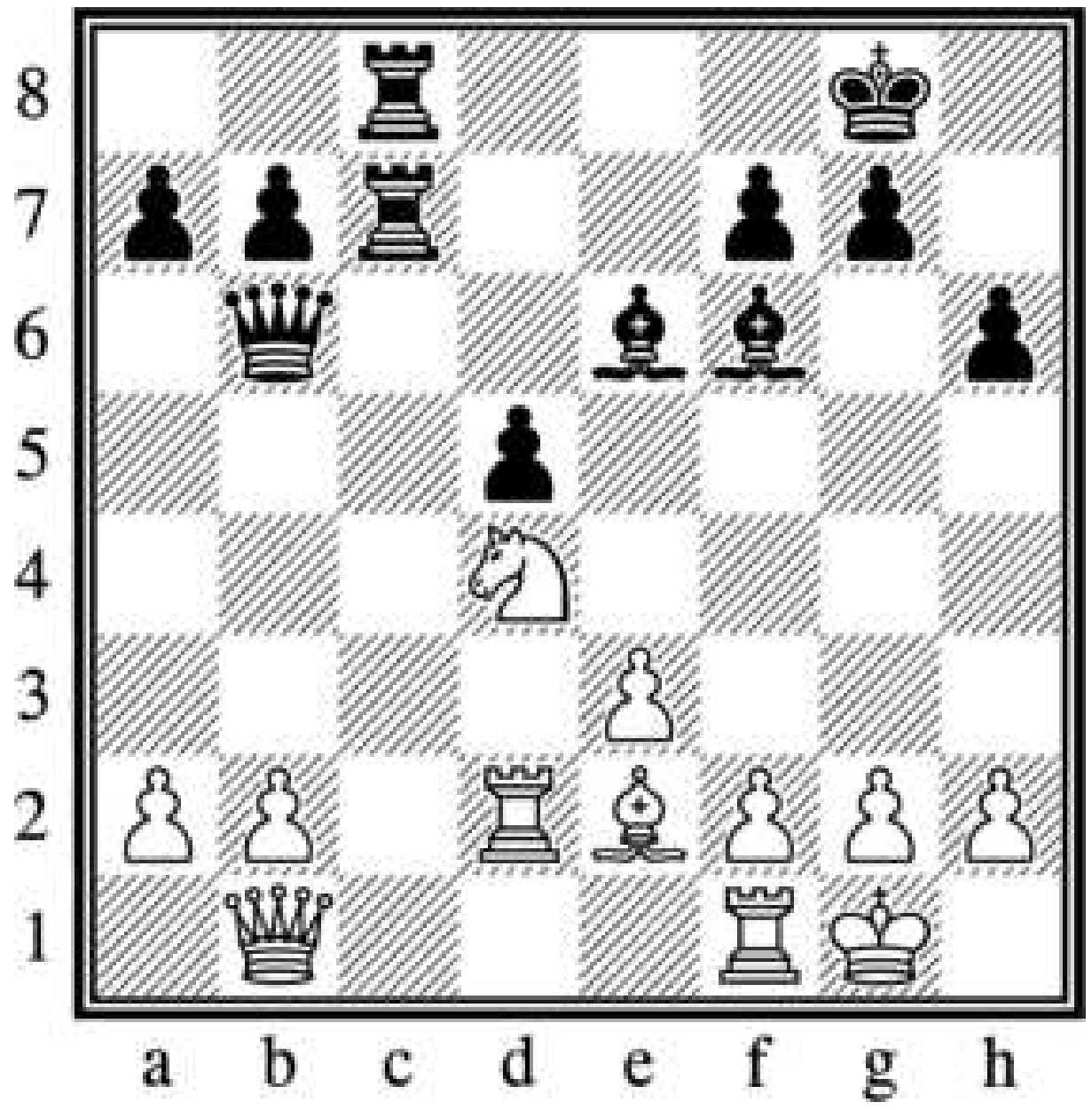

FIG. 1. *Karpov vs Kasparov key position, prior to White's 21st move:* 21. $N \times$ e6.

## 2. KARPOV–KASPAROV 1985: MOVE RUNS

Fischer has frequently claimed that all games of the 1985 WCC match between Karpov and Kasparov were rigged and prearranged move by move (Fischer, 1996; Polgár and Shutzman, 1997). Notably, at his first press conference following his recent release to Iceland from immigration detention in Japan, Fischer repeated this allegation (ESPN Sportscenter, March 25, 2005). One element of this accusation, indeed the only concrete piece of evidence proffered, concerns the fourth game and the position following Black's 20th move as depicted in Figure 1.

Fischer (1996) asserts:

> Starting on move 21, White makes no less than 18 consecutive moves on the light squares. Incredible!

Kasparov and Karpov played a total of 144 games that totaled 5540 moves over the course of their 5 WCCs—we will investigate this incredibility in terms of *runs*.

Consider $n$ i.i.d. Bernoulli trials with success probability $p$. Let $N_{n,k}$ be the number of nonoverlapping success runs of length $k$ and let $L_n$ be the length of the longest success run. Although our primary interest is in probabilities such as $\Pr(L_n > k)$, we will sometimes obtain these by exploiting the equivalence $\{L_n < k\} \Leftrightarrow \{N_{n,k} = 0\}$. The exact probability distributions for these events have been historically difficult to compute. Godbole (1990a) established the formula

$$\Pr(N_{n,k} = x)$$

$$(1) \quad = \sum_{\lfloor (n-kx)/k \rfloor \le y \le n-kx} q^y p^{n-y} \binom{y+x}{x} \cdot \sum_{0 \le j \le \lfloor (n-kx-y)/k \rfloor} (-1)^j \binom{y+1}{j} \binom{n-kx-jk}{y}.$$

While (1) offers computational advantages when contrasted with previously established combinatorially derived alternatives (see, e.g., Philippou and Makri, 1986), it is nonetheless problematic for evaluating the probability of observing a run of length $k = 18$ in a series of length $n = 5540$. However, Feller (1968), exploiting the theory of recurrent events, provides an approximation that is highly accurate even for moderate $n$,

$$(2) \qquad \Pr(L_n < k) \approx \frac{1 - p\theta}{k + 1 - k\theta} \cdot \frac{1}{\theta^{n+1}},$$

where $\theta$ solves $\theta = 1 + qp^k\theta^{k+1}$ and $q = 1 - p$. Applying (2) to the Karpov–Kasparov collection gives

$$\Pr(L_{5540} \ge 18) \approx 0.0105.$$

Now, this $p$-value is not incredibly incredible, especially since it does not accommodate Fischer's implicit search for other (run) patterns—presumably a run of moves to dark squares would also have registered. On the other hand, it allows runs to straddle different games and this does not reflect Fischer's likely ascertainment process. Rather, runs would be detected within games and, accordingly, individual games should constitute the units of analysis.

So, focusing on the key game, we are interested in $\Pr(L_{63} \ge 18)$, which can be evaluated using either (1) or (2). These give (with agreement to six decimal places)

$$(3) \qquad \Pr(L_{63} \ge 18) = 0.0000896.$$

If this $p$-value is adjusted for multiplicity (number of games = 144) via Bonferroni correction, we obtain an (adjusted) $p$-value (0.013) that again does not impress as being overly incredible. However, not only is such Bonferroni correction conservative, but it also does not reflect the length structure of the game collection. At the cost of obtaining $p$-values for the maximal run within each game, we could seek to redress these concerns using, say, step-down Bonferroni (Dudoit, Shaffer and Boldrick, 2003) or false discovery rates (Storey, Taylor and Siegmund, 2004). But there is a far more fundamental concern that affects the computation of these individual $p$-values. The results given in (1) and (2) require that the underlying sequence of Bernoulli trials is i.i.d. Here the corresponding $i$th random variable can be designated as $X_i \equiv I\{\text{White's move}_i \to \text{light square}\}$ and identically distributed equates to $p_i = \Pr(X_i = 1) = p \, \forall i$. All the above $p$-value calculations have been based on the seemingly natural specification $p_i = 0.5$. We next further scrutinize this specification.

The prescription $p_i = 0.5$ derives from the fact that 32 of the 64 squares on the chessboard are light. Indeed, the symmetries and piece and pawn moves are such that, prior to White's first move, there are equal numbers of moves to light and dark squares. However, that situation does not necessarily persist. For example, following Fischer's generally preferred first move, 1. e4, we have $p_2 \geq 2/3$. In studying hitting streaks in baseball, Albright (1993) employed so-called situational covariates to capture variation in underlying probabilities of getting a hit for a given at bat. These included such features as home field and day/night game. We could attempt an analogous approach here. Relevant covariates would include presence of opposite-colored bishops, presence of knights, nature of the pawn structure and even advent of a time control, the latter often inducing repetitions. However, selecting, characterizing and quantifying such positional covariates seems problematic, especially considering the availability of a simple proxy.

The proxy is based on the set of *legal* moves. After all, 18 consecutive light square moves in checkers

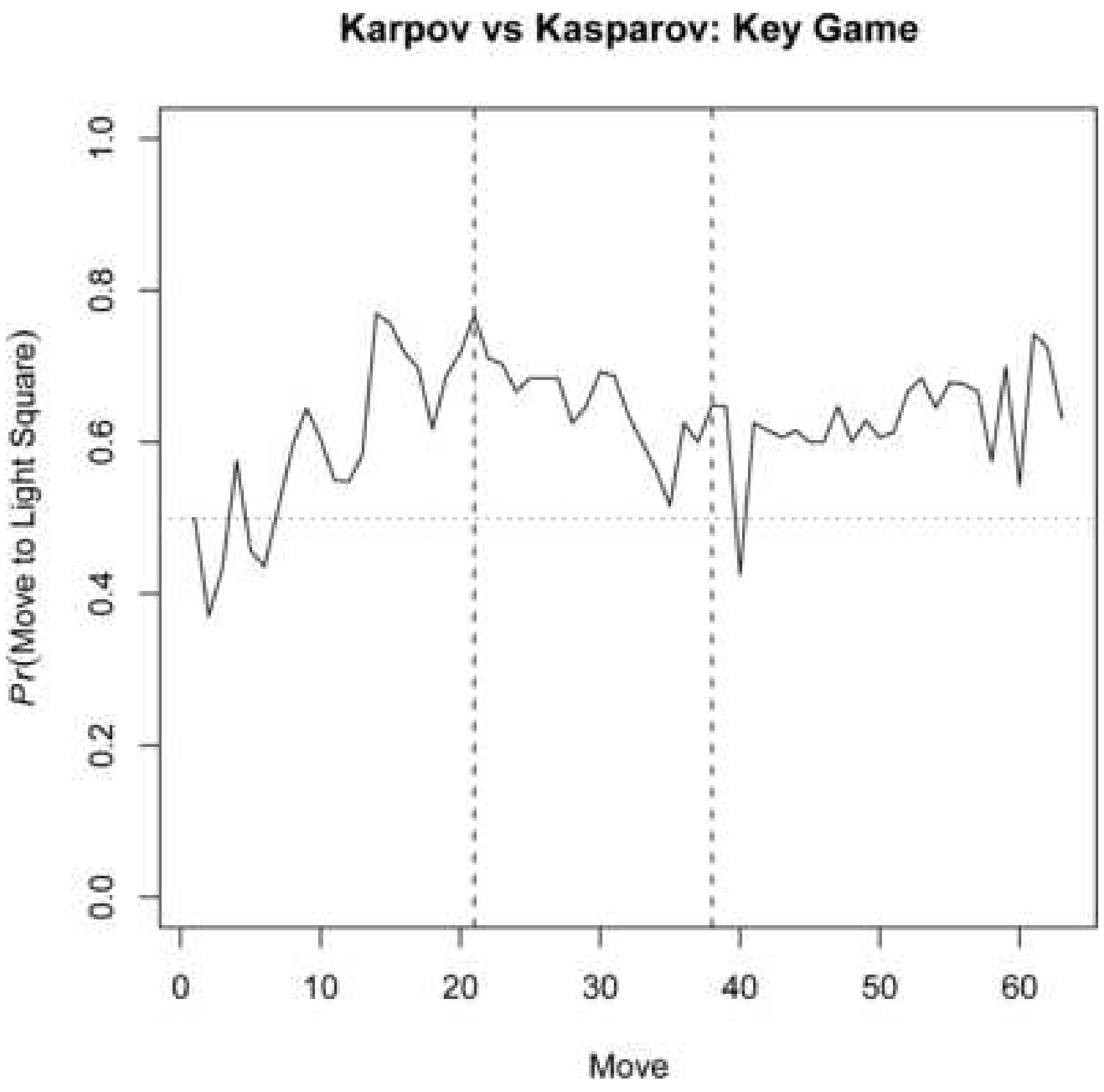

FIG. 2. *Probabilities of White making a* legal *move to a light square. The* 18 *move run is indicated by the dashed vertical lines.*

would be truly incredible, since checkers is played exclusively on dark squares. Define

$$p_i = \frac{\#\{\text{legal moves}_i \to \text{light square}\}}{\#\{\text{legal moves}_i\}}. \qquad (4)$$

Results that follow are not affected by how possible ambiguities surrounding castling are handled. Here, we exclude castling throughout. At the onset of the run we have $p_{21} = 33/43$. Using this value in either (1) or (2) gives

$$\Pr(L_{63} \geq 18 | p = p_{21}) = 0.096.$$

While again this is not remarkable, even prior to multiplicity adjustment, we have imposed $p_{21}$ throughout the game. Inspection of Figure 2, which plots $p_i$ vs $i$, reveals this to be an extreme choice. What is needed is a method for evaluating runs with varying success probabilities.

Fu and Koutras (1994) provided just such a methodology. They showed that the distributions of a variety of run statistics can be readily evaluated in the nonidentically distributed case by appropriate imbedding in a finite Markov chain. We briefly recapitulate the relevant definitions and results. For given $n$, let $\Gamma_n = \{0, 1, \ldots, n\}$ be an index set and let $\Omega = \{a_1, \ldots, a_m\}$ be a finite state space. A nonnegative, integer-valued random variable $Z_n$ can be imbedded into a finite Markov chain if:

(a) there exists a finite Markov chain $\{Y_i : i \in \Gamma_n = \{0, \ldots, n\}\}$ defined on $\Omega$,

(b) there exists a finite partition $\{C_x, x = 0, 1, \ldots, l\}$ on $\Omega$, and

(c) for every $x = 0, 1, \ldots, l$, we have $\Pr(Z_n = x) = \Pr(Y_n \in C_x)$.

If $Z_n$ can be imbedded into a finite Markov chain then, simply as a consequence of the definition and Chapman–Kolmogorov equations, we have (Theorem 2.1, Fu and Koutras, 1994)

$$\Pr(Z_n = x) = \pi_0 \left( \prod_{i=1}^{n} \Lambda_i \right) U'(C_x), \qquad (5)$$

where $\Lambda_i$ is the $m \times m$ transition probability matrix of the Markov chain, $\pi_0$ is the initial distribution and $U(C_x) = \sum_{r : a_r \in C_x} \mathbf{e}_r$ for $\mathbf{e}_r$ a $1 \times m$ unit vector having 1 at the $r$th coordinate and 0 elsewhere.

The art in utilizing imbedded Markov chains for evaluating run statistic distributions lies in eliciting suitable $\Omega$ and $\Lambda_i$. For our run statistic of interest, $N_{n,k}$ (recall equivalence to $L_n$), Fu and Koutras established $\Pr(N_{n,k} = 0) = \pi_0^* (\prod_{i=1}^n \Lambda_i^*) U^{*'}$, where

$$\underset{(k+1)\times(k+1)}{\Lambda_i^*} = \begin{bmatrix} q_i & p_i & 0 & 0 & \ldots & 0 \\ q_i & 0 & p_i & 0 & \ldots & 0 \\ q_i & 0 & 0 & p_i & \ldots & 0 \\ \vdots & \vdots & \vdots & \vdots & \ddots & \vdots \\ q_i & 0 & 0 & 0 & \ldots & p_i \\ 0 & 0 & 0 & 0 & \ldots & 1 \end{bmatrix}$$

and $\pi_0^* = (1, 0, \ldots, 0)$ and $U^* = (1, \ldots, 1, 0)$ are $1 \times (k+1)$ vectors. For the key Karpov–Kasparov game use of this imbedding, with $p_i$ as per (4) and Figure 2, gives

$$\Pr(L_{63} \geq 18 | p_i) = 0.0065. \qquad (6)$$

The densities (smoothed histograms) for $L_{63}$ under $p_i$ and $p_i = p = 0.5$ are presented in Figure 3. The shift in mass corresponding to (appropriate) use of legal moves is evident and underscores the difference between the $p$-values (3) and (6). We note that computation via imbedded Markov chains is exceedingly fast and easy.

Since (6) is our final $p$-value proffered for Karpov and Kasparov's 18-move run of light square moves, we provide some context with regard to magnitude by citing Short and Wasserman's (1989) appraisal of the $p$-value they computed in assessing the "streak-of-streaks"—Joe DiMaggio's 56-game hitting streak. They reported one $p$-value of 0.0000486 and comment that "the paucity of this probability is not overwhelming." Accordingly, we would *not* characterize the $p$-value in (6) as incredible.

## 3. CHESS PROGRAMS: FRITZ

In baseball, the batters objective is to hit, the 1919 Chicago Black Sox excepted. In basketball, where streak shooting has also been statistically evaluated (Larkey, Smith and Kadane, 1989; Tversky and Gilovich, 1989a, b), the objective is to make the bucket, with 1978–1979 Boston College (among others) excepted. In chess, if the objective was to move to light squares, there would be a lot more sacrifice of dark square bishops. So, to infer anything about "signal" or "suspicion" in the moves that constitute the run singled out by Fischer, it is necessary to judge the quality of these moves. If, conditioning on successive positions within the streak, the best move happens to be to a light square, then what is the message/implication? Is it incredible if the World Champion makes 18 good moves in a row? Was the quality of moves played during the run notably different than the quality of moves chosen outside the run? Of course, judging quality is tricky—Fischer, Karpov and Kasparov are all former World Champions and regarded among the greatest players of all time—it would be presumptuous for mere mortals to second guess their evaluations.

So, to make such judgments we turn to chess-playing software. Select programs are now capable of playing competitively against the world's leading grandmasters. For example, in October 2002, World Champion Vladimir Kramnik drew an eight-game match against the Fritz program; in early 2003, Kasparov drew a six-game match against the Deep Junior program; and in November 2003, Kasparov drew a four-game match against Fritz. Accordingly, we use Fritz to evaluate move quality. While we contend that these results, along with wider tournament performances and ratings, establish Fritz's credentials to determine good moves, it is important to recognize that we are not claiming infallibility or even superiority of Fritz's move assessments. Rather, since comparisons will be relative—{in run $(R)$/not in run $(\bar{R})$}—all that is required is that strong moves are consistently recognized and that the assessments are unbiased with respect to the move run.

Fritz returns a quantitative score for each move evaluated. Here, the more positive the score, the better the move. We cannot just compare $R$ and $\bar{R}$ scores directly, since differing score distributions reflect the decisiveness of the move and so, for example, are correlated with stage of the game. Thus, for our first outcome, we standardize according to the best move possible and define $\Delta_i = \text{score}_i - \text{bestscore}_i$, where $\text{score}_i$ is Fritz's score for the $i$th move as played and $\text{bestscore}_i$ is the score for the best possible move. So, $\Delta_i \leq 0$ with less negative values corresponding to better moves. Comparing the $\Delta$s obtained during the run $(n_R = 18)$ to those not in the run $(n_{\bar{R}} = 45)$ via a two-sample $t$-test shows that better moves tended to be played during the run, but the difference is nonsignificant $(p = 0.27)$. The same findings hold if we limit nonrun moves to the nine preceding and nine succeeding the run (to exert finer control over stage of game), and/or if we work on a relative scale ($\Delta_i^* = \Delta_i/\text{bestscore}_i$), and/or if we work with ranks instead of scores.

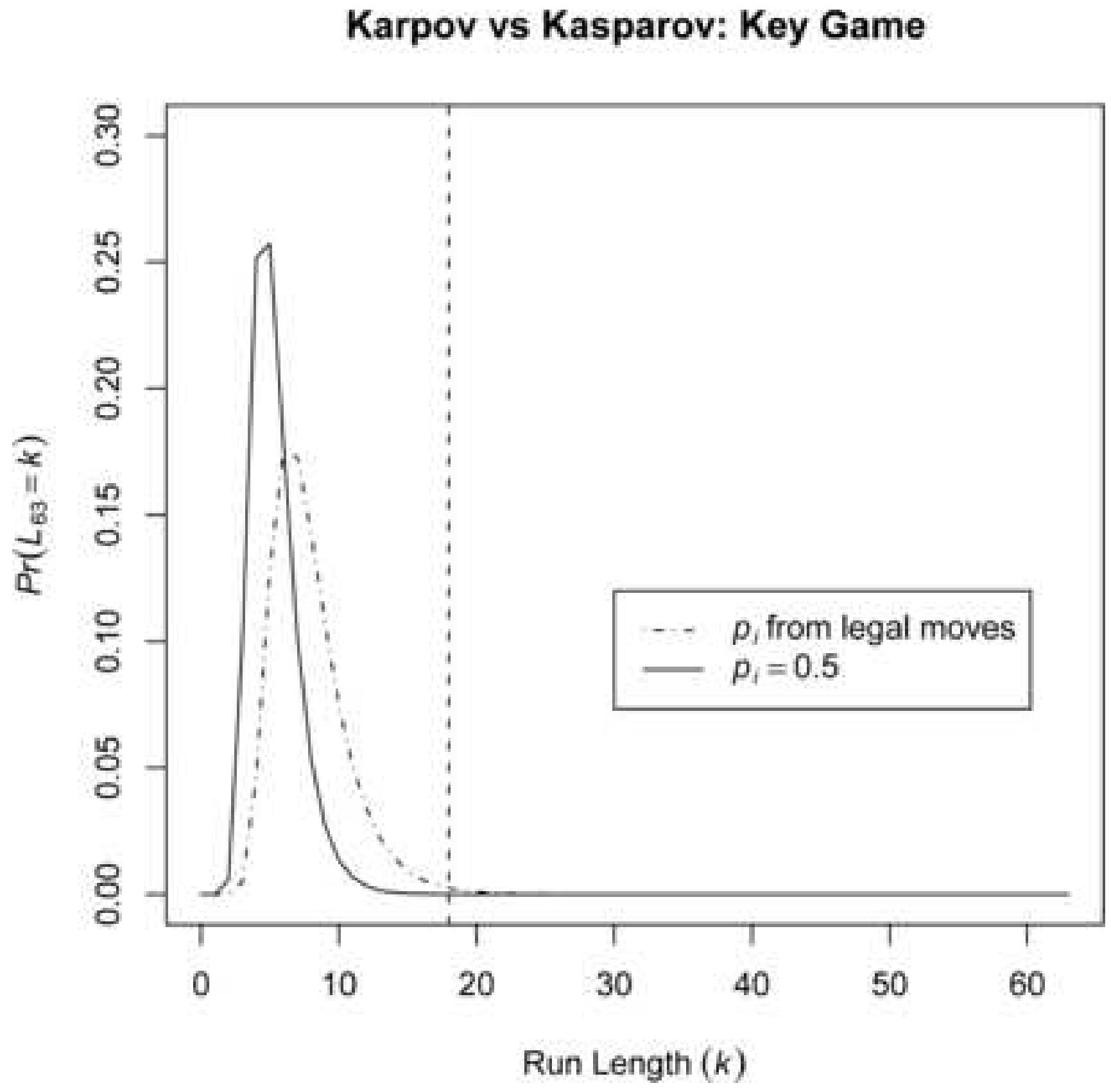

FIG. 3. *Smoothed histograms of* $\Pr(L_{63} = k)$. *Tail areas to the right of the dashed line at* $k = 18$ *correspond to the p-value reported in* (3) *and* (6).

A more sophisticated approach is afforded by employing the binary outcome that indicates whether the move played coincided with Fritz's best move, $Y_i = I\{\Delta_i = 0\}$, and using logistic regression to attempt to adjust for putative covariates. This also remedies a concern with $t$-test appropriateness in view of the underlying mixed (discrete with mass at zero, continuous) distribution of $\Delta$. As previously, covariate specification is challenging: what we propose here is natural but not exhaustive. Playing the best move when it is obvious is less incredible than when it is obscure. Accordingly, we adjust for positional complexity which, for each move, we operationalize as $\text{complexity}(j,k) = \text{score}(j\text{th best})/\text{score}(k\text{th best})$ for differing choices of $j > k = 1,\dots,5$. Similarly, playing the best move when the number of possibilities is limited is also less compelling, so we use $\text{possibles} = \#\{\text{legal moves}\}$ as another covariate. We fitted several models of the form $\text{logit}(Y) = \beta\text{run} + f(\text{possibles}, \text{complexity}(j,k))$, corresponding to different specifications for $f, j$ and $k$. All gave highly null results with respect to tests for $\beta$, the parameter of interest.

Therefore, when considering the quality of moves played, there is nothing distinctive, let alone incredible, about the run Fischer isolates. However, in addition to relying on Fritz's assessments of move quality, these analyses could be criticized on the basis of being underpowered to detect distinguishing attributes of the run. So, as a final approach, we turn to repositories of chess games to provide broader context for the run.

## 4. SEARCHING CHESS GAME DATA BASES

Efron (1971) reexamined Bode's law "governing" mean planetary distances to the Sun as an exemplar of whether an observed sequence of numbers follows a simple rule. He states that differing analytic approaches would have been employed had measurements on 50 solar systems been available. So far, our treatment of the significance of the run has been confined to the key Karpov–Kasparov game in question. However, unlike the planetary system situation, there are extensive data bases of chess games that we can mine so as to appraise the distinctiveness of the run identified by Fischer. To this end we conducted three structured searches.

The first was based on the key position itself (1). If this position had occurred in other high level games, then the ensuing sequence of moves would be potentially informative with regard to the incredulity of the run. Not surprisingly, search of an online data base that comprises more than two million games (`www.chesslab.com/`) revealed the position to be unique.

The next search is motivated by an important feature of the key position: the presence of opposite-colored bishops. This is arguably the most important attribute with respect to generating move runs on either light or dark squares. We used the commercial Chessbase (`www.chessbase.com/`) Big 2000 data base which, although not as sizable as Chesslab (1,327,059 games), possesses more powerful search tools. The following search parameters were prescribed—for a game to be selected it must satisfy all requirements at some stage:

1. Opposite-colored bishops
2. No knights or major pieces
3. Three to four pawns
4. Average Elo rating of players $\geq 2500$
5. Length of game $\geq 80$ moves

In addition to limiting the number of games chosen, these criteria are motivated by the following considerations. Item 2 eliminates pieces that can move on either color squares; item 3 balances complexity; item 4 imposes grandmaster level play. Item 5, which mandates relatively long games, is motivated by run statistic asymptotics. In particular, if $n, k \to \infty$ such that $nqp^k \to \lambda$, then $\Pr(N_{n,k} = x)$ tends to the Poisson probability $e^{-\lambda}\lambda^x/x!$ (Feller, 1968; Godbole, 1990b). So if $p$ is fixed, $k = O(\log(n))$, thereby bestowing a premium on examining longer games.

The resulting search of Chessbase Big 2000 yielded 11 games that had, in chronological order, the following maximal run lengths: $L_{85} = 21, L_{109} = 29, L_{81} = 46, L_{90} = 10, L_{92} = 13, L_{85} = 18, L_{81} = 17, L_{86} = 14, L_{98} = 20, L_{113} = 9, L_{94} = 12$. Fischer's identified Karpov–Kasparov run, with $L_{63} = 18$ does not stand out against this collection. Indeed, if anything is notable, it is the third game with $L_{81} = 46$: not only was this game played at the candidates level (Timman vs Salov, 1988), but, in addition to the cited 46-move run (played by Black) that attains an appreciably smaller $p$-value than the Karpov–Kasparov run ($2.24 \times 10^{-13}$), it featured a separate 34-move run (played by White).

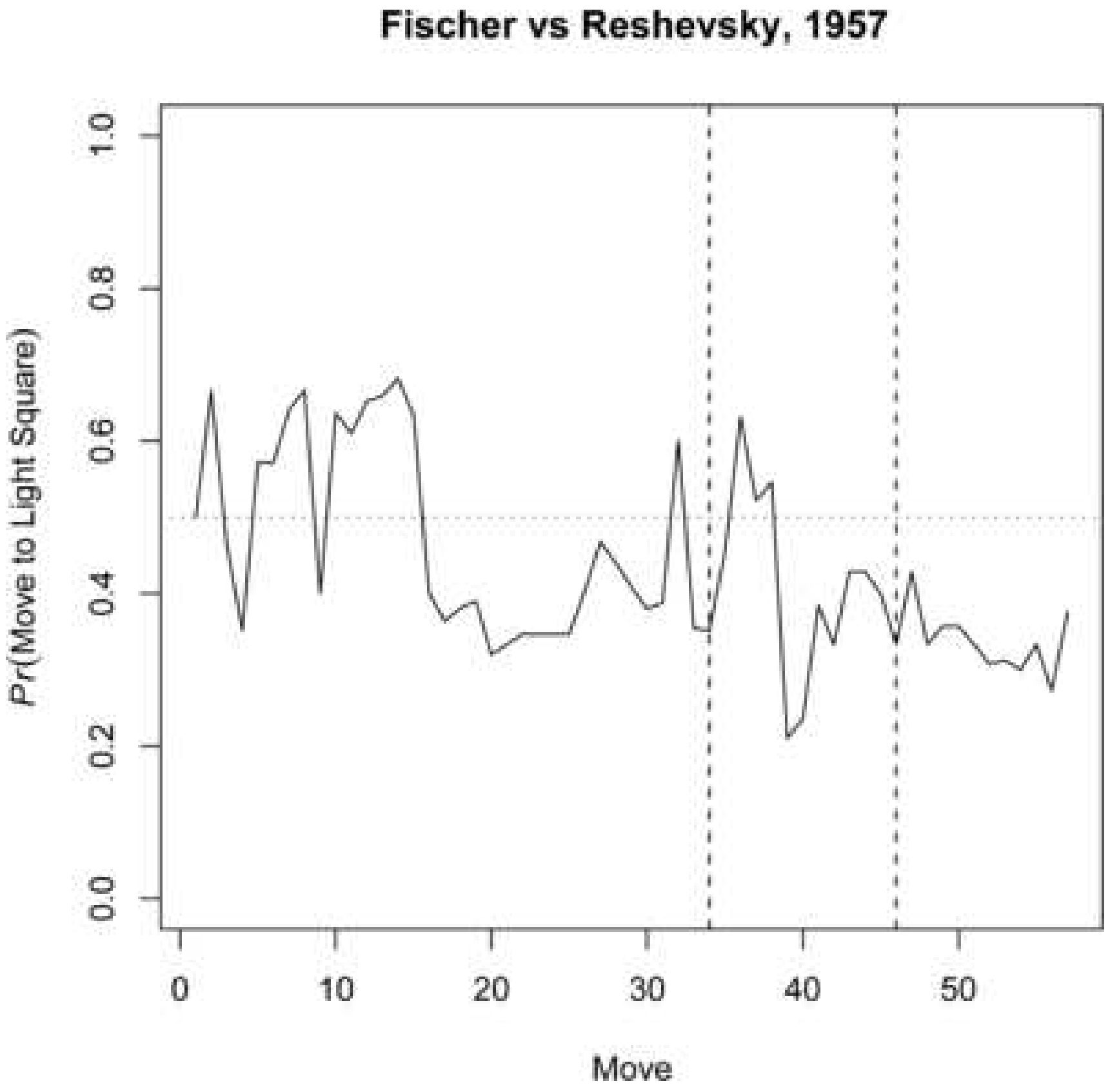

Fig. 5. *Probabilities of White making a legal move to a light square. The 13-move run is indicated by the dashed vertical lines.*

For the third and final search we elected to scrutinize the 827 Chessbase Big 2000 games of Fischer himself. Search item 1 was retained, the restrictions imposed by items 2, 3 and 4 were eliminated, and item 5 was relaxed to game lengths $\geq$ 50 moves. This yielded a total of five games. We focus on one of these, Fischer vs Reshevsky, 1957, United States Championship. In particular, the position arising prior to White's 34th move is of interest; see Figure 4. From this position Fischer (White) played 13 consecutive moves to dark squares, this in the context of a 57-move game with move-by-move probabilities of moving to a light square as depicted in Figure 5.

We follow the same approach in using imbedded Markov chains to evaluate the significance of this run, using $p_i$ in accordance with (4) and Figure 5, as was employed in Section 2. This gives

$$\Pr(L_{57} \geq 13 | p_i) = 0.0023. \tag{7}$$

The punch line is that the $p$-value in (7) is more extreme than that achieved by the Karpov–Kasparov run given in (6). Thus, if the latter is "incredible," what does this imply about the former? It is of interest to note how the $p_i$ fluctuations in the Fischer–Reshevsky game, as depicted in Figure 5, are such that the probability density for $\Pr(L_{57} = k)$ under $p_i$ from (4) is barely distinguishable from that under $p_i = 0.5$, as illustrated by the smoothed histograms in Figure 6.

## 5. DISCUSSION

In their evaluation of coincidences, Diaconis and Mosteller (1989) showcased the roles played by misperception, multiplicity and the law of truly large numbers. These factors have arguably contributed to Fischer's overstating the significance of the Karpov–Kasparov move run. Indeed, misperceptions surrounding the significance of runs arising from random processes are purportedly commonplace; see, for example, Scheaffer, Gnanadesikan, Watkins and Witmer (1996) for pedagogic exercises that illustrate this point. These misperceptions are compounded when (implicit) search for an extreme run is performed and multiplicity considerations are ignored; see Al-bert (2004) for some examples from baseball. In addition to seemingly failing to account for these concerns when assessing run significance, Fischer, in expressing incredulity at the Karpov–Kasparov move run, has also assumed a constant "success" probability for each move: $p_i = p = 0.5$.

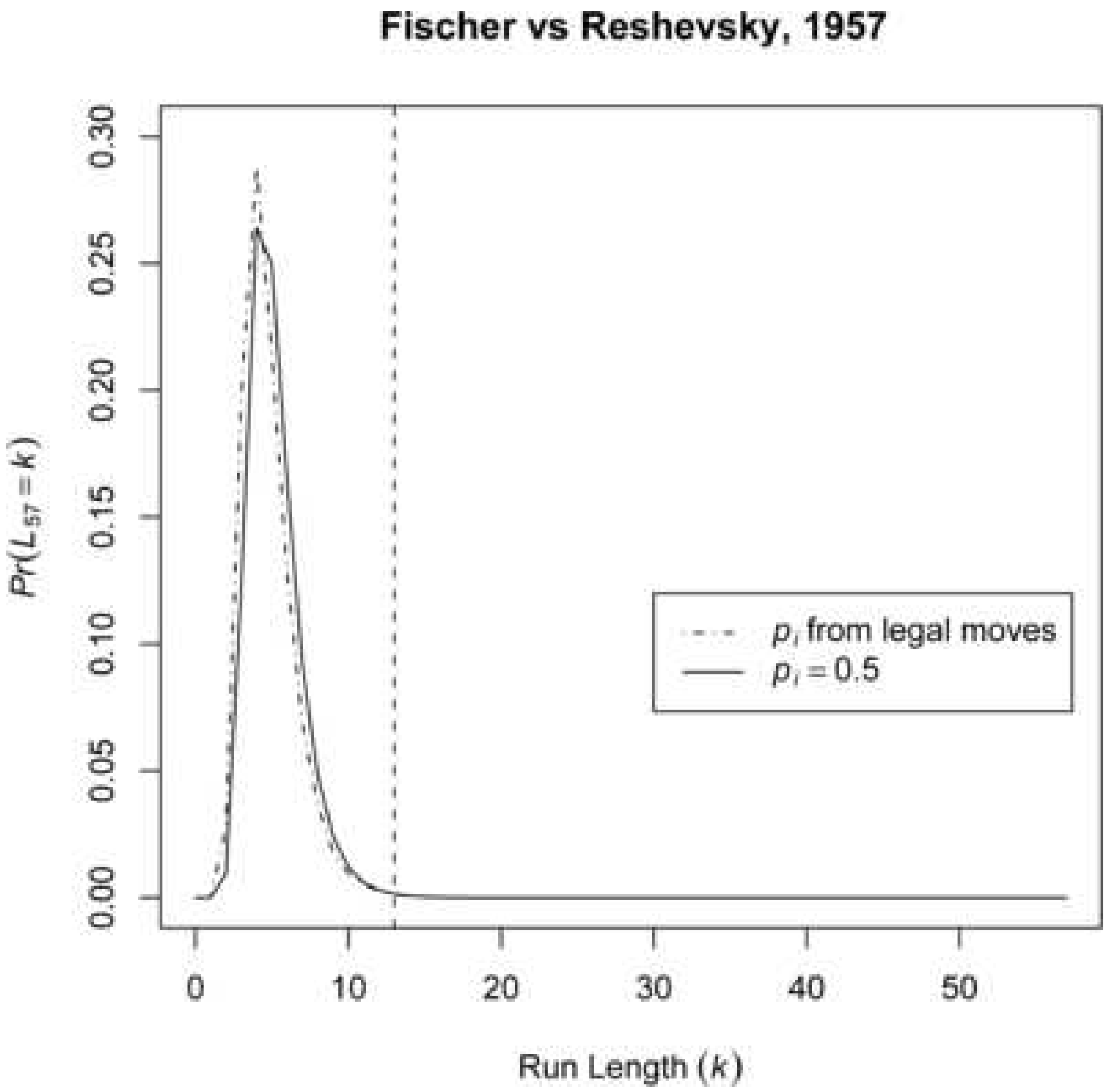

FIG. 6. *Smoothed histograms of* $\Pr(L_{57} = k)$.

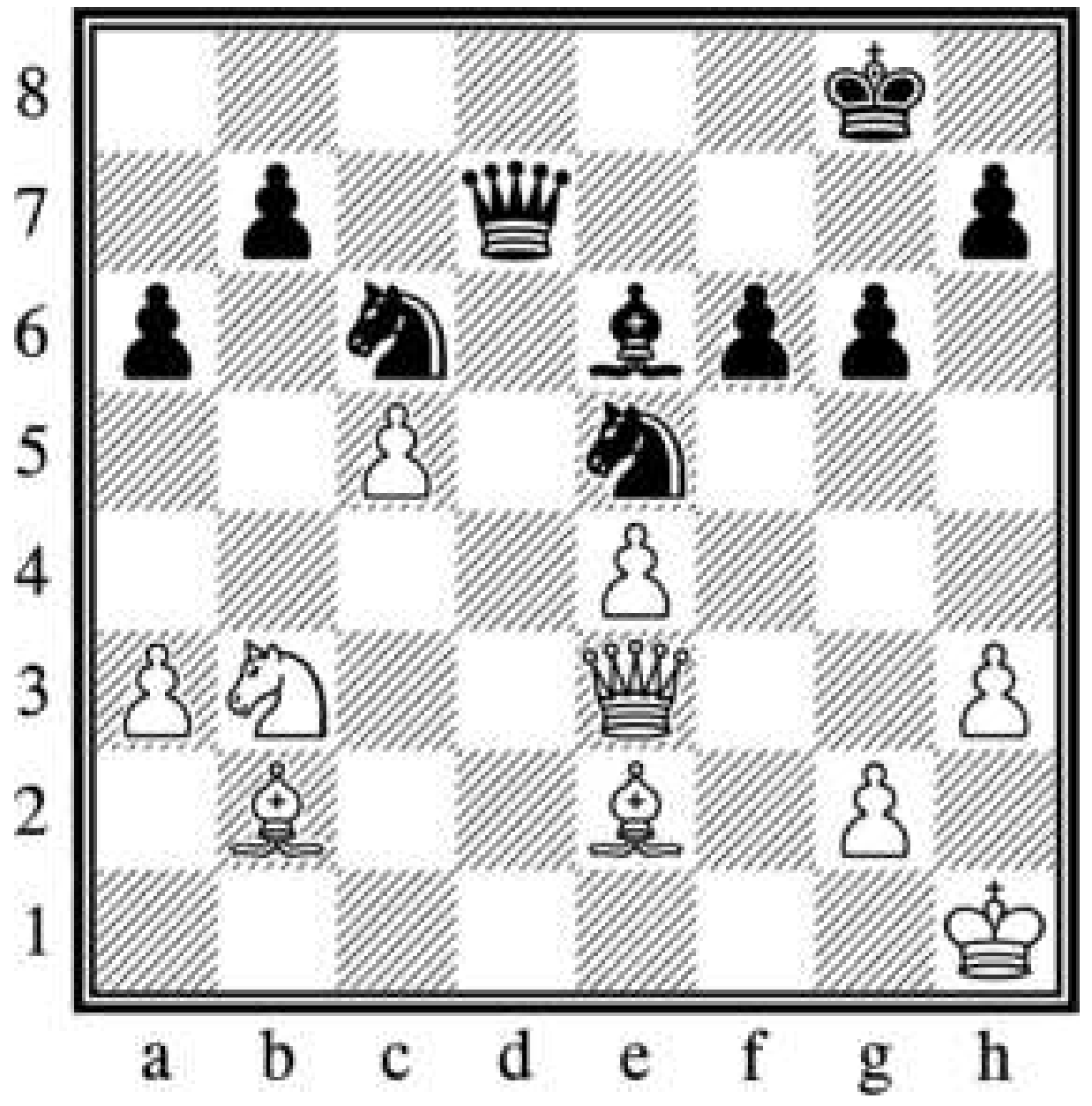

FIG. 4. *Fischer vs Reshevsky key position, prior to White's* 34th *move:* 34. *N*d4.

Even aside from variation in $p_i$, it is clear from (1) or (2) that $\Pr(L_n \geq k)$ depends heavily on $p$. This has again been noted in the baseball context (Casella and Berger, 1994)—a more successful player (team) is likely to have a longer hitting (winning) streak than a less successful player (team). Indeed, Lou (1996) provided an elegant extension to the imbedded Markov chain approach of Fu and Koutras (1994) that allows for evaluation of joint and conditional distributions of the number of successes and $L_n$. Not only does her framework accommodate varying $p_i$, but it also facilitates testing of between-move dependency. However, application of these methods here is problematic owing to difficulties in estimating, or even prescribing, the requisite transition probabilities. Furthermore, the power curves displayed by Lou (1996) suggest that in our $n = 63$ setting we will be hard-pressed to detect between-move dependencies. Additionally, as noted by Rubin, McCulloch and Shapiro (1990), there are instances where such conditional analyses are inappropriate. These include situations when there is no intrinsic interest in the total number of successes, as is the case here where success corresponds to a move to a light square.

Alternative approaches to appraising the significance and/or asserting the existence of runs and streaks have been advanced. In the sports context, Yang (2004) employs Bayesian binary segmentation to assess whether transitions in success probability are evident. Applying his methodology to the key Karpov–Kasparov game yields highly null results.

As highlighted by a referee, evaluation of success runs can be highly dependent on (i) the sample space employed and (ii) how the success probabilities, $p_i$, are framed. Here, we have argued that (i) the appropriate unit of analysis is a game, rather than a match or series of matches, and (ii) allowing variable $p_i$, that at the least reflects the number of legal moves available, is essential. The resultant achieved significance levels were sufficiently null [see (6)] that the above mentioned multiplicity corrections [for the number of games in the Karpov–Kasparov match(es)] were not pursued.

However, in other settings, including two celebrated examples briefly described next, these considerations can be less clear-cut. In their evaluations of Joe DiMaggio's 56-game hitting streak (baseball), Short and Wasserstein (1989) performed various probability calculations that are readily reproduced using either (1) or (2). What distinguishes the differing calculations are the alternative sample spaces considered: season, career, history of baseball. Indeed, dramatically different results ensue, with no attempt to arbitrate or reconcile between them. Throughout, they employ constant $p$ based on lifetime batting average. In this case, given the rich documentation surrounding baseball, it might be possible to improve on this imposition. This could make recourse either to day-of-game batting average or, more ambitiously, to modeling $p_i$ based on game specific covariates, akin to Albright (1993). However, as is made clear by the discussants of that article, such modeling is problematic. Furthermore, despite the extensive compilation of historical baseball statistics, extracting the requisite game level data is, at best, a laborious undertaking.

Tversky and Gilovich (1989a) contended that the "hot hand" phenomenon in basketball is a "cognitive illusion" that derives from misconceptions of the laws of chance. They base this assertion on analyses of data obtained by coding shot attempts from 48 televised NBA games plus free throw sessions of college players. In an effort at refutation, Larkey, Smith and Kadane (1989) took issue with the analyses in large part on the basis of sample space considerations. They argued that the unit of analysis should reflect "cognitively manageable chunks," capturing temporal proximity, which they operationalize as 20 consecutive shot attempts. Tversky and Gilovich (1989b) offer a rebuttal that (in part) further refines sample space considerations by accommodating individual playing times. Here a definitive characterization of sample space seems daunting. More challenging still is capturing variations in $p_i$, despite the widely acknowledged need to do so. This derives from the need to accommodate the hard-to-quantify notion of defensive pressure, above and beyond facets such as position on the floor and phase of the game. Further compounding these obstacles is the absence of usable data bases: the analyses conducted required the individual investigators to watch and encode film, which in and of itself led to dramatic disputation (Tversky and Gilovich, 1989b).

In our evaluations of the significance of the Karpov–Kasparov move run, not only have we benefited from being able to frame the problem (relatively) precisely, but also from the existence of sophisticated chess-playing software and game data bases. The latter tool could be used to assess a run potentially far more remarkable than the move run, namely, Fischer's own aforementioned run of 19 consecutive wins as part of the 1972 WCC qualifying cycle. This run included two 6–0 shutouts in his two candidates matches. As an indication of the incredibility of these results and to ground things in the world of grounders, *Time* magazine equated the shutouts with pitching "two straight no-hitters." Perhaps Fischer's ascent to World Champion was part of some conspiracy.

## ACKNOWLEDGMENTS

I thank the Executive Editor, an Editor, two referees and Chuck McCulloch for many helpful comments and suggestions.